\documentstyle[prl,aps,preprint]{revtex}
\def\thru#1{\mathrel{\mathop{#1\!\!\!/}}}
\begin{document}
\tighten

\title{THE NUCLEON'S TENSOR CHARGE\thanks
{This work is supported in part by funds provided by the U.S.
Department of Energy (D.O.E.) under cooperative agreement
\#DF-FC02-94ER40818.}}

\author{Hanxin He\thanks{Permanent address:
Institute of Atomic Energy, Beijing, China} and Xiangdong Ji}

\address{Center for Theoretical Physics \\
Laboratory for Nuclear Science \\
and Department of Physics \\
Massachusetts Institute of Technology \\
Cambridge, Massachusetts 02139 \\
{~}}

\date{MIT-CTP-2380 ~~~ HEP-PH/9412235 ~~~
Submitted to: {\it Phys. Rev. D1} ~~~ November 1994}

\maketitle

\begin{abstract}
We seek to understand the physical significance of the nucleon's
tensor charge and make estimates of its size in phenomenological
models and the QCD sum rule.
\end{abstract}

\pacs{xxxxxx}

\narrowtext

The nucleon's {\it tensor charge} $\delta \psi$ ($\psi=u, d, s, \ldots$)
is defined as the forward matrix element of the
tensor current, $T^{\mu\nu} = \bar \psi \sigma^{\mu\nu}\psi$,
in the nucleon state,
\begin{equation}
    \langle PS|\bar \psi\sigma^{\mu\nu}\psi|PS
      \rangle = \delta \psi  \,\bar U(PS)\sigma^{\mu\nu}U(PS) \>,
\label{def}
\end{equation}
where $P$ is the nucleon's four-momentum, $S$ is a
polarization vector, and $U(PS)$ is a Dirac spinor.
Due to the $\gamma$-matrix identity,
$ \sigma^{\mu\nu}\gamma_5 = (i/2)\epsilon^{\mu\nu\alpha\beta}
\sigma_{\alpha\beta} $, one can also define the tensor charge
in terms of the operator $\bar \psi \sigma^{\mu\nu}i\gamma_5\psi$,
and then the right hand side of Eq.~(\ref{def}) becomes $2\delta \psi
(P^\mu S^\nu-P^\nu S^{\mu})$. Throughout the paper, we
adopt the notations in Itzykson and Zuber~\cite{IZ}.

Like other nucleon charges ({\it baryon charge} defined by
the matrix element of $\bar
\psi\gamma^\mu\psi$, {\it axial charge} by $\bar
\psi\gamma^\mu\gamma_5\psi$, and {\it scalar charge} by $\bar
\psi\psi$), the tensor charge is one of the fundamental
parameters that characterize properties of the
nucleon. So far, however, little is known about
its value and its implication on the structure of the
nucleon. In this paper we seek to understand
the physical significance of the tensor charge
and make estimates in the MIT bag model and
the QCD sum rule.

The main reason for lack of studies about
the tensor charge is that it is difficult to
access experimentally. There are no fundamental
probes which couple directly to the tensor current.
[Before the V-A weak interaction was firmly established,
physicists had entertained the possibility of
weak scalar and tensor couplings.]
However, the situation has changed fundamentally
when the factorization theorems in high-energy
scattering are shown to be valid on quite general
ground~\cite{CSS}. The theorems provide
a firm basis for the general parton-model result that
the perturbative scattering in
hard processes effectively provides
a versatile probe into structure of
hadrons. One recent example of such application
is the measurement of the
nucleon's axial charge from polarized lepton-nucleon
scattering~\cite{SMC}.

It was first discussed by Ralston and Soper \cite{RS}
that the transversely-polarized Drell-Yan scattering
can probe a new quark distribution of the nucleon,
the transversity distribution $h_1(x)$.
What is the $h_1(x)$ distribution?
Consider a nucleon traveling in $z$-direction
with its polarization in $x$-direction.
The polarization of
quarks and antiquarks in the nucleon
can be classified in term of
the transversity eigenstates $|\uparrow\downarrow\rangle
=(|+\rangle \pm |-\rangle)/\sqrt{2}$, where $|\pm\rangle$
are the usual helicity eigenstates. If one uses
$N_\uparrow(x)$ ($N_\downarrow(x)$) to represent the density of quarks
with polarization $|\uparrow\rangle$($|\downarrow\rangle$), then
\begin{equation}
h_1(x) = N_\uparrow(x) - N_\downarrow (x) \>,
\end{equation}
and likewise for antiquarks.  The $h_1(x)$, together with
the unpolarized quark distribution $q(x)$ and the quark
helicity distribution $g_1(x)$ forms a complete set
for describing the quark state inside the nucleon in
the leading-order hard processes. It was demonstrated by
Jaffe and Ji \cite{JJ} that the first moment of $h_1(x)$ is related to
the nucleon's tensor charge,
\begin{equation}
     \int^1_{-1} h_1(x) dx = \int^1_0(h_1(x)-\bar h_1(x))dx =
      \delta \psi
\label{sum}
\end{equation}
where $h_1(x)$ at negative $x$ is negative of the
antiquark distribution $\bar h_1(-x)$. Given no
fundamental tensor coupling, the integral may be the
best hope to gain knowledge about the tensor charge.
In Ref.~\cite{JI},
other possible experiments of measuring the transversity
distribution are examined. The RHIC spin collaboration and the HERMES
collaboration have proposed a first measurement of $h_1(x)$
in the future~\cite{RHHE}.

According to Eq.~(\ref{sum}), the nucleon's tensor charge
measures the {\it net number of trans\-versely-pol\-arized
valence quarks} (quarks minus antiquarks) in a transversely-polarized
nucleon. One would argue that this number should be
the same as the net number of longitudinally-polarized
valence quarks in a longitudinally
polarized nucleons (which is related to the axial charge),
since, after all, a polarization
of the nucleon in its rest frame can be said longitudinal,
or transverse, or combinations of both.
This argument would be correct if the nucleon
were made of free quarks.
During high-energy scattering,
quarks in the nucleon do appear to be free. However,
rotational invariance now becomes non-trivial
because high-energy processes
select a special direction. In fact,
in the so-called
infinite momentum frame where parton model was originally
formulated, the rotational
operators explicitly involve interactions~\cite{DRA}. Thus the
difference between the tensor and axial
charges has a dynamical origin.

Unlike the baryon or axial charges, the tensor charge is
renormalization-scale dependent. A simple calculation of the
anomalous dimension for $\bar \psi\sigma^{\mu\nu}\psi$ yields,
\begin{equation}
     \gamma = 2C_F {g^2\over 16\pi^2} + \cdots,
\end{equation}
where $C_F = 4/3$, $g$ is the strong coupling constant,
and dots represent higher-order terms in the coupling. Thus
$\delta \psi$ scales according to,
\begin{equation}
     \delta \psi(\mu^2) =
       \left( {\alpha(\mu^2)\over \alpha(\mu_0^2)} \right)^{4\over 33-2n_f}
 \delta \psi(\mu_0^2) \>,
\end{equation}
where $n_f$ is the number of flavors. As $\mu^2\rightarrow \infty$,
$\delta \psi$ vanishes. This contrasts the nucleon's
scalar charge $\langle P|\bar \psi\psi|P\rangle$, which
scales as $\left({\alpha(\mu^2)/\alpha(\mu_0^2)}\right)^{-{12\over 33-2n_f}}$,
and blows up as $\mu^2\rightarrow \infty$.

Now, let us consider the size of the tensor charge in
non-relativistic quark models~\cite{JJ}. In the limit
of $m_q \rightarrow \infty$, the
transverse-spin operator commutes with a free-quark hamiltonian
and so the transverse polarized quarks are in the
transverse-spin eigenstates. Then rotational
invariance implies,
\begin{equation}
     \delta \psi = \Delta \psi \>,
\end{equation}
where $\Delta \psi$ is a conventional notation for the axial
charge. Or,
\begin{eqnarray}
   \delta u&  = &{4\over 3} \>, \nonumber \\
           \delta d &=&-{1\over 3} \>, \\
               \delta s& =& 0 \>.\nonumber
\end{eqnarray}
This result can also be obtained from the fact that
the tensor operator $\bar \psi \sigma^{0i}i\gamma_5\psi$
differs from the axial current $\bar \psi \gamma^i\gamma_5\psi$
by a $\gamma_0$ factor, which reduces to 1 in the non-relativistic
limit.

In the MIT bag model, the tensor charge
can be expressed in terms of the upper
and lower components ($f$ and $g$) of the quark wave function~\cite{JJ},
\begin{eqnarray}
          \delta u & = & {4\over 3}\int (f^2+{1\over 3}g^2) \>,  \nonumber \\
           \delta d & = & -{1\over 3}\int (f^2+{1\over 3}g^2) \>.
\end{eqnarray}
This differs from the expressions for the
nucleon's axial charge by a sign in front of $g^2$,
due to the same $\gamma_0$ factor mentioned above.
Instead of trying to find the best bag parameters, we
demand $\Delta u -\Delta d =1.257$ and use
the normalization $\int(f^2+g^2)=1$, then the
tensor charge is uniquely fixed,
\begin{eqnarray}
           \delta u & =& 1.17 \>,\nonumber \\
           \delta d& =& -0.29 \>.
\end{eqnarray}
These numbers are closer to the non-relativistic quark
model result than are the nucleon's axial charge in the bag.
In another words, the non-relativistic quark model
prediction for the tensor charge appears to be
less susceptible to relativistic effects than
for the axial charge.

Of course, these estimates in phenomenological
models are very crude and
provide only a guidance at the best. In particular,
the matching between QCD quarks and constituent quarks
used in models is a subtle and unsolved problem.
This is reflected by the fact that
model calculations have no explicit
reference to any scale, although one would generally believe
that these models live in a scale somewhere in between
$\Lambda_{\rm QCD}$ and the nucleon mass.
More reliable estimates can be made
with QCD-based approaches in which one deals with QCD
quarks directly. One approach is the lattice
QCD. The recent progress in calculating
axial and scalar charges on lattice shows that
the lattice QCD becomes increasingly competitive
with other methods in computing hadron observables~\cite{KFL}.
Another approach is the QCD sum rule.
In the past fifteen years, this method
has produced a large number of interesting
results which are largely consistent with hadron
phenomenology~\cite{SHI}. In the remainder of this paper,
we present a QCD sum rule estimate of the tensor charge.

There exist in the literature several equivalent
formulations of the QCD sum rule technique for calculating
forward hadron matrix elements. Following the approach
initiated by Balitsky et al.~\cite{BY}, we consider the
following three-point correlation function,
\begin{equation}
W^{\mu\nu} =i^2 \int d^4xd^4y e^{ip\cdot x}
    \langle 0|T[\psi\sigma^{\mu\nu}\psi(y)\eta(x)\bar\eta(0)]|0 \rangle \>,
\end{equation}
where $\eta$ is the nucleon interpolating field,
$\eta = \epsilon^{abc}u^T_aC\gamma_\mu u_b \gamma_5\gamma^{\mu}d_c $,
and $C=i\gamma^2\gamma^0$ is the charge conjugation matrix.
We calculate $W^{\mu\nu}$ at large Euclidean $-p^2$
using the operator-product-expansion
technique on the one hand, and using resonance saturation on the
other. The tensor charge is extracted by matching the two
results at certain kinematic domain where both methods
are supposed to be valid.

In resonance saturation, $W^{\mu\nu}$
contains the nucleon double pole, single pole,
and other resonance contributions,
\begin{eqnarray}
    W^{\mu\nu} &= &\delta \psi \left(\thru p\sigma^{\mu\nu}\thru p
      + m_N^2\sigma^{\mu\nu} + m_N\{\thru p, \sigma^{\mu\nu}\} \right)
     {\lambda^2 \over (p^2-m_N^2)^2} + \cdots \nonumber \\
         & = & W_1\thru p\sigma^{\mu\nu}\thru p
           + W_2\sigma^{\mu\nu}  + W_3\{\thru p, \sigma^{\mu\nu}\}
              + \cdots
\label{pole}
\end{eqnarray}
Here we have shown only the double pole term, in which
$\lambda^2$ is the coupling of the nucleon with the interpolating
field,
$\langle 0|\eta(0)|p\rangle = \lambda U(p)$. Other terms
are neglected because they either vanish or are
suppressed after multiplied by $(m^2_N-p^2)$
and the Borel transformation.
There are three different Dirac structures emerging
from the double-pole term: chiral-odd ones with coefficients
$W_1$ and $W_2$ and chiral-even one with coefficient $W_3$,
each of which can be used to construct a sum rule
and extract $\delta \psi$.
In principle one has to obtain the same result
from each of them before one trust the final
answer. Or else one can obtain any desired result by making
different combinations of the sum rules.
In practice, however, some sum rules are better
approximated by leading power corrections than others.
Thus, choosing a right sum rule to extract the physical
observable is a very delicate issue.

Depending upon momentum flow in Feynman diagrams, the
operator-product expansion for $W^{\mu\nu}$ has
three distinct classes of contributions: perturbative,
local and bi-local power corrections. The perturbative contribution
comes from large momentum flow through {\it all} internal lines
of diagrams. The local-power contribution refers to diagrams in
which some particle lines are condensed into vacuum and
the momentum flowing through the composite
operator $\bar \psi \sigma^{\mu\nu} \psi$ is large. The bi-local
power contribution is similar to the local one
except the momentum flowing through the
composite operator is infrared. To evaluate the bi-local
contribution, one needs two-point correlation
functions at zero momentum,
\begin{equation}
      \int d^4 x \langle 0|T \left[O_n(0)\bar \psi \sigma^{\mu\nu}
          \psi(x)\right] |0\rangle \>,
\end{equation}
where $O_n$ are local operators from the operator-product expansion of
$T\eta(x) \bar \eta(0) = \sum_n C_n(x^2) O_n(0)$.  These
two-point functions can be evaluated either in terms
of the QCD sum rule, or, in some cases, with QCD equations
of motion. The bi-local contribution is similar in spirit
to the contribution from vacuum susceptibility
introduced by Ioffe et al.~\cite{Ioffe}.

Now we present the sum rule results of the tensor
charges for the up and down quarks separately.
For the $u$ quark, the leading large-momentum
contribution to $W_1$ and $W_2$
comes from the power corrections
with a dimension-six condensate,
\begin{eqnarray}
 W_1 &=& {2\over p^4} \langle \bar u u \rangle
 \langle\bar dd \rangle+ \cdots\nonumber  \\
 W_2 &=& - {2\over 3p^2} \langle \bar u u \rangle
\langle\bar dd \rangle   + \cdots,
\end{eqnarray}
where $W_1$ receives contribution from
both local and bi-local power terms, whereas
$W_2$ from a bi-local power term alone.
Following the standard procedure of
multiplying by $m_N^2-p^2$, making
Borel-transformation, and matching with the
corresponding term from
Eq.~(\ref{pole}), we find for the $W_1$ sum rule,
\begin{equation}
       \delta u = {2\over \lambda^2} \langle \bar uu \rangle^2
   e^{m_N^2/M^2} \>.
\end{equation}
As the Borel mass $M^2$ changes from $m_N^2$ to $2m_N^2$, $\delta u$
changes by about fifty percent, so the sum rule is reasonably
stable. Taking $\langle\bar uu\rangle = -(240{\rm MeV})^3$, $M^2=m_N^2$,
$\lambda^2=7.0\times 10^{-4}{\rm GeV}^6$, we get,
\begin{equation}
         \delta u = 1.0 \sim 1.5 \>.
\end{equation}
On the other hand, the result from the $W_2$ sum rule is
smaller by a factor of three.
Without calculating higher-order terms, it is difficult
to determine which one is more reliable. However, experiences with
other sum rules indicate that the result from $W_1$
with non-vanishing local contribution is more stable
against higher-order corrections.

The contribution to the chiral-even $W_3$ comes from the
dimension-three and five power corrections,
\begin{equation}
   W_3 = {1\over 2\pi^2} \ln(-p^2) \langle \bar uu\rangle
           + {1\over 24\pi^2p^2} \ln({-p^2\over \mu^2}) \langle \bar ugG\cdot
              \sigma u\rangle  + \cdots
\end{equation}
where $\mu^2$ is a infrared cut-off that can be taken to be $\Lambda_{\rm
QCD}^2$. After the Borel transformation, we get
at $M^2=m_N^2$,
\begin{equation}
   \delta u = -{m_Ns_0 \over 2\pi^2\lambda^2}
              e^{1-s_0/m_N^2} \langle \bar uu\rangle
            - {em_0^2m_N \over 24 \pi^2\lambda^2}
             (\ln({m_N^2\over \mu^2})-1)\langle\bar uu\rangle \>.
\end{equation}
Taking $s_0 = (1.5{\rm GeV})^2$ and $m_0^2 = 0.8 {\rm GeV}^2$, we get,
\begin{equation}
      \delta u = 0.94 \>.
\end{equation}
Combining the above results, we conclude that
the leading-order sum rule calculation gives
\begin{equation}
       \delta u = 1.0 \pm 0.5
\end{equation}
at the scale of $\mu^2=m_N^2$.

Next, we consider the $d$-quark tensor charge.
Due to its chiral-even property, $W_3$
receives local power corrections only from odd-dimensional
condensates. A simple consideration
shows that such contributions start
with the dimension-nine condensate,
$\langle \bar \psi \psi\rangle^3$. This suggests that $d$ quark
tensor charge is quite small.
The suspicion is confirmed by the consideration of
other two chiral-odd sum rules.

For $W_2$, the leading contribution comes from
a perturbative term, followed by a power
correction associated with the dimension-four condensate
$\langle {\alpha_s\over \pi} G^2\rangle $.
Neglecting the latter, we have,
\begin{equation}
   W_2 = {1\over 32\pi^4} p^4\ln -p^2  + \cdots ,
\end{equation}
which yields,
\begin{equation}
  \delta d =  0.3    \>,
\end{equation}
at $M^2 = m_N^2$, a number indeed quite small.
The leading contribution to $W_1$ comes from a
bi-local correlator $\Pi(0)$,
\begin{equation}
   W_1 = - {1\over 144\pi^2} \ln(-p^2) \Pi(0) +\ldots,
\end{equation}
where $\Pi(0)$ is,
\begin{equation}
      \Pi(0) = i\int d^4x \langle 0|
           T[\bar d\sigma^{\alpha\beta}d(0)
           \bar d\sigma_{\alpha\beta}d(x) ]|0\rangle \>.
\end{equation}
Assuming $\rho(1^{--})$ and $B (1^{+-})$
meson dominance, and estimating the relevant
coupling constants in the
QCD sum rule, we find
\begin{equation}
           \Pi(0) \sim (0.15 {\rm GeV})^2
\end{equation}
The small $\Pi(0)$ result comes from the cancellation of
the two resonances, and thus the theoretical error
on the estimate is large. The $W_1$ sum rule produces,
\begin{equation}
       \delta d = {\Pi(0)\over 144\pi^2\lambda^2}M^2(m_N^2-M^2)
                  e^{m_N^2/ M^2} \,.
\end{equation}
Depending upon of a choice of the Borel parameter,
$\delta d$ is in the range of 0.0 to $-0.1$,
Given uncertainties with different sum
rules, we conclude,
\begin{equation}
         \delta d = 0.0 \pm 0.5
\end{equation}
at $\mu^2 = m_N^2$.
This is consistent with a recent
QCD sum rule calculation for the transversity
distribution $h_1(x)$~\cite{IK}.

To recapitulate, the leading-order QCD sum rule
suggests $\delta u = 1.0 \pm 0.5$ and $\delta d=0.0 \pm 0.5$
at the scale of about 1 GeV$^2$.
A recent SU(3)-symmetric, leading-order
large-$N_c$ analysis \cite{JO} shows that $\delta u + \delta d$
is on the order of $1/N_c$ relative to $\delta u - \delta d$.
This result on the flavor structure also applies to
the axial charge, for which an analysis of a recent measurement
\cite{SMC} yields
$\Delta u = 0.78$ and $\Delta d = -0.46$, a favorable
comparison with the large $N_c$.
If the true value of $\delta d$ is indeed rather small as
the QCD sum rule indicates, the large $N_c$ analysis
perhaps has little relevance for the tensor charge.

In summary, we discussed in this paper various aspects
of the nucleon's tensor charge. We focused on its numerical
value in the MIT bag model and the QCD sum rule.
With various caveats, the both results seem consistent.
Admittedly, the QCD sum rule calculation is done only
at the leading order, one must show that the results
are stable against higher-order power corrections
and that all sum rules for the same quantity
yield same answer. Nonetheless, we believe
our result is qualitatively reliable.
Clearly, a lattice QCD calculation or a direct experimental
measurement of the tensor charge will produce
a more definitive determination of this interesting
observable.

We thank I. Balitsky for numerous
discussions on QCD sum rule calculation. H. X. He is
grateful to the Center for Theoretical Physics at MIT
for the warm hospitality during his visit.

\end{document}